\documentclass[12pt,a4paper]{article}
\usepackage{t1enc}
\usepackage[dvips,final]{graphicx} 
\usepackage{tabularx} 
\usepackage{amsfonts}
\usepackage{amsmath}

\def\du{\unskip\smash{\lower 1.4ex \hbox{\char34}}\kern-.2ex}
\def\hu{\kern-.2ex\hbox{\char92}}

\newcommand{\bdis}{\begin{displaymath}}
\newcommand{\edis}{\end{displaymath}}
\newcommand{\be}{\begin{equation}}
\newcommand{\ee}{\end{equation}}

\newcommand{\mcal}{\mathcal}
\newcommand{\pd}{\partial}

\newcommand{\mE}{\langle E\rangle}
\newcommand{\Z}{\mathcal{Z}}

\begin{document}
\baselineskip=6.5mm
\newpage

\title{Particle in a box at high temperature}
\author{Michal Demetrian\footnote{{\it
demetrian@fmph.uniba.sk}}  \qquad \\
Comenius University,
Mlynska Dolina M 105, \\ 842 48 Bratislava IV,
Slovakia}
\maketitle

\abstract{High temperature expansion of the partition function for a
particle on a segment of a line is found to show an
example of the quantum system that thermodynamical functions do not
approach the thermodynamical functions of its classical counterpart at
high temperature. The problem might be interesting for the students and
teachers because by a collection of noninteracting
particles trapped in a box at sufficiently high temperature and low density
one imitates the classical ideal gas.}

\section{Introduction}

This section briefly reminds few basic facts that can be found
in almost any textbook of statistical physics or solid
state physics, see for instance refs. \cite{hu}, \cite{ll},
\cite{reif}.
Let us start with the classical linear harmonic oscillator which dynamics
is given by the Hamiltonian
\be \label{1.1}
\mcal{H}=\frac{1}{2m}p^2+\frac{1}{2}m\omega^2x^2 \quad ,
\ee
where $m$ is the mass, $\omega$ is the angular frequency and coordinates
$p$ and $x$ run from $-\infty$ to $\infty$. The mean energy $\mE^C$ and
specific heat $C^C$ of
the oscillator are given by the Boltzmann distribution
\be \label{1.2}
\mE^C=\frac{\int_{-\infty}^\infty{\rm d}p\int_{-\infty}^\infty{\rm d}x
e^{-\beta \mcal{H}}\mcal{H}}
{\int_{-\infty}^\infty{\rm d}p\int_{-\infty}^\infty{\rm d}xe^{-\beta \mcal{H}}}=kT,
\quad C^C=\frac{{\rm d}\mE^C}{{\rm d}T}=k \quad ,
\ee
where we have introduced the inverse temperature $\beta=1/kT$. The above
written result is nothing but the special case of the equipartition
theorem and Dulong-Petit law known from the theory of specific heat of
solids. \\
Now, let us turn to the case of quantum linear oscillator, for simplicity,
in one dimension. Energy levels
($E_n=\hbar\omega(1/2+n), \quad n\in\{0,1,2,\dots \}$)
of this system and its eigenfuctions
can be found in any textbook of quantum
mechanics. It is easy to compute the partition function
$\Z=\sum_{n=0}^\infty\exp(-\beta E_n)$ and then one can compute the mean
energy of the oscillator $\mE^Q$ and its specific heat $C^Q$.
The results read
\be \label{ehq}
\mE^Q=\frac{1}{2}\frac{\hbar\omega}
{\tanh\left(\frac{1}{2}\hbar\omega\beta\right)}, \quad
C^Q=\frac{1}{4}k\beta^2\frac{(\hbar\omega)^2}
{\sinh^2\left(\frac{1}{2}\hbar\omega\beta\right)} \quad .
\ee
By performing the high temperature ($\beta\hbar\omega\ll 1$)
expansion of the formulas one obtains the results
(\ref{1.2}). In the following section we will show that this relation
between classical and quantum statistics of a physical system is not only
possible.


\section{Particle on an interval within classical and quantum statistical
physics}

Let us consider the particle of mass $m$ that can move on the segment
of a line with the length $L$. The mean energy and the specific heat of such a particle at the temperature $T$ are given,
within classical statistical physics, by
\be \label{ce}
\mE^C=\frac{\int_{-\infty}^\infty{\rm d}p\int_0^L{\rm d}x
\exp\left[-\frac{p^2}{2m}\beta\right]\frac{p^2}{2m}}
{\int_{-\infty}^\infty{\rm d}p\int_0^L{\rm d}x
\exp\left[-\frac{p^2}{2m}\beta\right]}=\frac{1}{2\beta}=\frac{kT}{2}\ , \quad
C^C=\frac{{\rm d}\mE^C}{{\rm d}T}=\frac{k}{2} \quad
.
\ee

\begin{figure}[h]
\centering
\includegraphics[width=8cm,height=6cm]{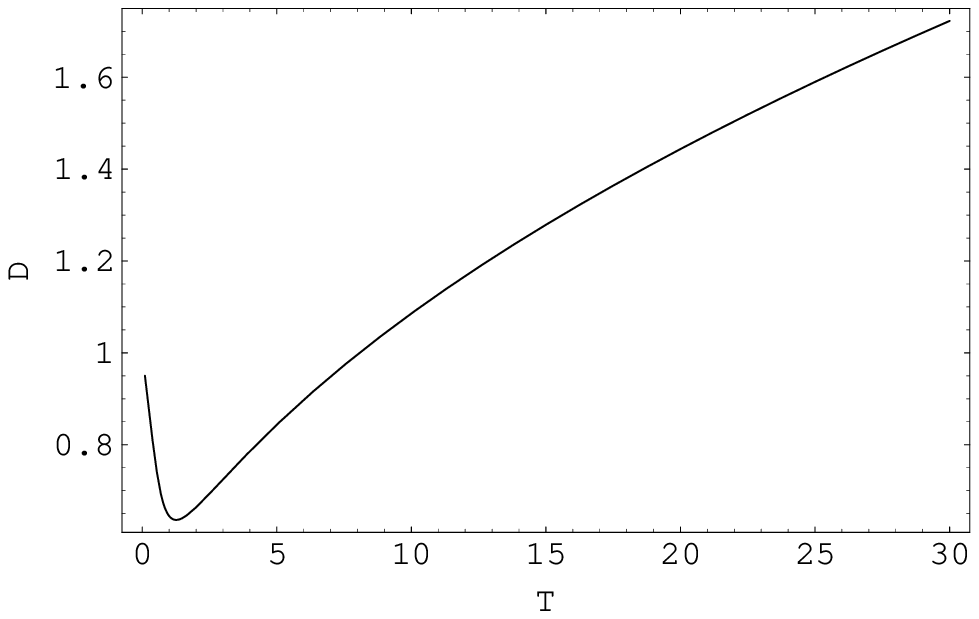}
\includegraphics[width=8cm,height=6cm]{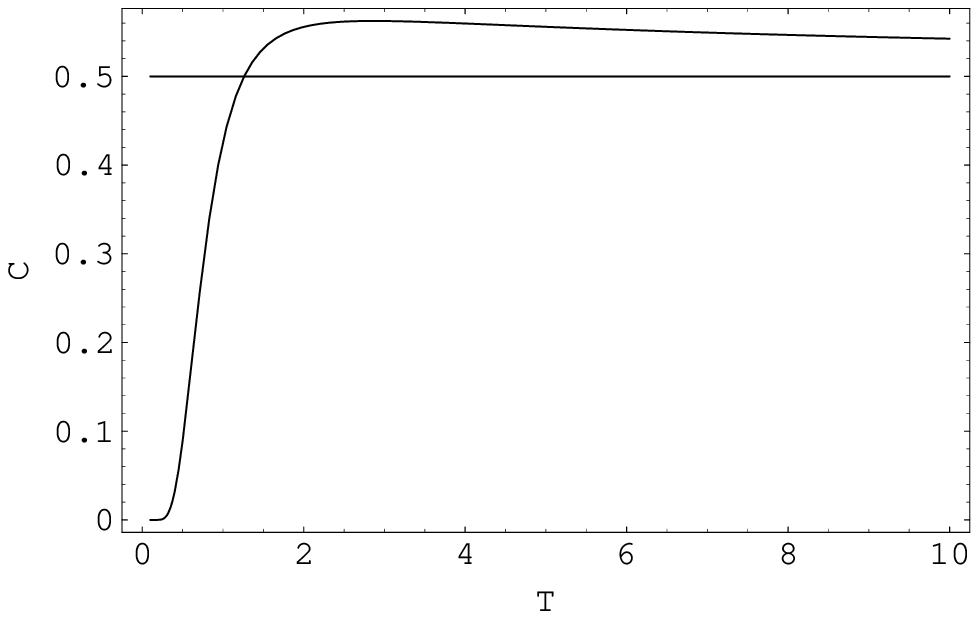}
\caption{Left graph shows the temperature
dependence of the difference between the mean energy
given by the eq. (\ref{mefp}) and equipartition term $T/2$. On the right
graph we see the temperature dependence of the specific heat (\ref{c1}).
We
see, that specific heat approaches the classical value $1/2$ as $T$ goes
to infinity. Specific heat vanishes at $T\to 0^+$ and there is one maximum
of $C$ which could be expected from the high-temperature expansion of
specific heat which follows from the eq. (\ref{mefp}).}
\label{f2}
\end{figure}

Within the quantum mechanics we should impose some boundary conditions on
the wave function $\psi=\psi(x), \quad x\in [0,L]$.
Let us suppose the following boundary conditions:
$\psi(0)=\psi(L)=0$ hold. Then the eigenenergies
of the particle are given by the equation
\be \label{esfp}
E_n=\frac{\pi^2\hbar^2}{2mL^2}n^2, \quad n\in\{ 1,2,3, \dots \} \quad .
\ee
We wish to compute the mean energy of this particle at the
temperature $T$ and to show that in the limit $T\to\infty$ mean energy
does not approach the equipartition value $kT/2$. For the simplicity,
we have chosen the
system of units and the mass of the particle
that the energy spectrum (\ref{esfp})
will read $E_n=n^2, \quad n\in\{ 1,2,3, \dots \}$ (the Boltzmann constant equals $1$).
The mean energy is given by
\be \label{e1}
\mE^Q=\frac{\sum_{n=1}^\infty n^2\exp[-n^2\beta]}
{\sum_{n=1}^\infty \exp[-n^2\beta]}=
-\frac{\frac{\pd }{\pd\beta}\sum_{n=1}^\infty \exp[-n^2\beta]}
{\sum_{n=1}^\infty \exp[-n^2\beta]} \quad ,
\ee
and the specific heat is given by
\be \label{c1}
C^Q=\frac{1}{T^2}
\frac{\left(\sum_{n=1}^\infty n^4e^{-n^2/T}\right)
\left(\sum_{n=1}^\infty e^{-n^2/T}\right)-
\left(\sum_{n=1}^\infty n^2e^{-n^2/T}\right)^2}
{\left(\sum_{n=1}^\infty e^{-n^2/T}\right)^2} \quad .
\ee
The above written sums, unfortunately, can not be evaluated in terms of
elementary functions. However,  the high-temperature expansion of these sums
can be found quite easily, as shown below. We deal with
\be \label{suma}
\Z(\beta)=\sum_{n=1}^\infty \exp[-n^2\beta] \quad .
\ee
To perform the high-temperature ($\beta\ll 1$) expansion of this sum we will
use the Abel - Plana formula:
\bdis
\sum_{n=0}^\infty F_n(x)=\frac{1}{2}F_0(x)+\int_0^\infty{\rm d}tF_t(x)+
i\int_0^\infty\frac{F_{it}(x)-F_{-it}(x)}{e^{2\pi t}-1}{\rm d}t \quad .
\edis
In our case we have
\begin{eqnarray*}
\Z(\beta) & = &
-\frac{1}{2}+\int_0^\infty e^{-x^2\beta}{\rm d}x+
i\int_0^\infty\frac{e^{-ix^2\beta}-e^{ix^2\beta}}
{e^{2\pi x}-1}{\rm d}x \\
& = & -\frac{1}{2}+\frac{1}{2}\sqrt{\frac{\pi}{\beta}}+
2\int_0^\infty\frac{\sin(\beta x^2)}{e^{2\pi x}-1}{\rm d}x \quad .
\end{eqnarray*}
We denote the last term in the above written equation by $I(\beta)$ and
we can have (for $\beta\ll 1$):
\bdis
I(\beta)\approx 2\beta\int_0^\infty\frac{x^2}{e^{2\pi x}-1}{\rm d}x=
\frac{2\beta}{(2\pi)^3}\sum_{k=1}^\infty\int_0^\infty{\rm d}z z^2
e^{-kz}=\frac{4\beta}{(2\pi)^3}\zeta(3) \quad ,
\edis
where $\zeta(x)=\sum_{k=1}^\infty 1/(k^x), \quad \mbox{Re}(x)>1$ is the Riemann
zeta function and $\zeta(3)\approx 1.20206$.
So, we have obtained the statistical sum $\Z(\beta)$ with the accuracy up
to the first order in $\beta$ in the form
\be \label{z}
\Z(\beta)=-\frac{1}{2}+\frac{1}{2}\sqrt{\frac{\pi}{\beta}}+
\frac{4\zeta(3)}{(2\pi)^3}\beta \quad .
\ee
Having the result (\ref{z}) it is easy to derive the mean energy, the result reads:
\be \label{mefp}
\mE^Q=-\frac{\pd \ln[\Z(\beta)]}{\pd\beta}=
\frac{T}{2}+\frac{1}{2\sqrt{\pi}}T^{1/2}+\frac{1}{2\pi}-\frac{3\zeta(3)}
{2\pi^{7/2}}T^{-1/2}+O(T^{-1}) \quad .
\ee
We see that the mean energy of our particle within quantum statistical
physics differs from the mean energy of such a particle within classical
statatistical physics (\ref{ce}) and this difference grows with the
temperature like $T^{+1/2}$ - so it is not true that in the limit
$T\to\infty$ one has to obtain classical results from the quantum ones.
To compare results (\ref{mefp}) and (\ref{ce}) one should rewrite the result
(\ref{mefp}) into standard
system of units in which the first two terms of
(\ref{mefp}) read
\bdis 
\mE^Q=
\frac{1}{2}kT+\left(\frac{\pi}{8}\right)^{1/2}\frac{\hbar}{\sqrt{m}L}
(kT)^{1/2}+O(1) \quad .
\edis
It is natural that the correction term (with respect to the equipartition term $kT/2$)
is proportional to the Planck constant. Planck
constant makes it small for the realistic values of $m$ and $L$. It would
be useful to realize that in the three dimensional analogue of our problem
we would get the following results: $\mE^C_{3D}=3\mE^C$ and
$\mE^Q_{3D}=3\mE^Q$. \\
On the other hand the specific heat approaches the equipartition value
$k/2$ at $T\to\infty$.
The temperature dependence of the specific heat for both low and high
temperatures is shown in figure \ref{f2}.

\subsection{Simple derivation of the three leading terms of the eq.
(\ref{mefp})}

We would like to show how to get the first three terms of eq.
(\ref{mefp}) in a elementary way. Our next
approach to the problem will be less accurate and will not offer the way how to get
next terms of the expansion (\ref{mefp}). We take the sum
(\ref{suma}) and perform the following manipulations:
\begin{eqnarray*}
\Z(\beta) & = &
\frac{1}{2}\left(\sum_{n=-\infty}^\infty e^{-n^2\beta}-1\right)
=\frac{1}{2}\left(\int_{-\infty}^\infty
{\rm d}xe^{-x^2\beta}+f(\beta)-1\right) \\
& = & \frac{1}{2}
\left[\left(\frac{\pi}{\beta}\right)^{1/2}+f(\beta)-1\right] \quad ,
\end{eqnarray*}
where we have defined the function $f$.
If we take for granted that the function $f$ and its first derivative
are sufficiently small at small $\beta$ we can deal with it as follows:
\bdis
\mE^Q=
\frac{\frac{1}{2}\left[\frac{1}{2}\sqrt{\pi}\beta^{-3/2}-f'(\beta)
\right]}{\frac{1}{2}\left[\sqrt{\pi}\beta^{-1/2}+f(\beta)-1\right]}
\approx
\frac{1}{2}T+\frac{1}{2\sqrt{\pi}}T^{1/2}+\frac{1}{2\pi} \quad .
\edis
Finally, we have obtained the part of the result from previous paragraph in a
relatively easy manner.

\section{Discussion and concluding remarks}

It can be very easy for a student to get used to believe that for any
quantum system the high temperature limits of its thermodynamical functions
have to be equal to their classical counterparts. We have shown in the
second
section of this work that this statement is not true. We have found the system for which
the absolute difference between the mean energy in the classical theory and quantum theory grows to infinity with
growing temperature. However, the relative difference between these quantities vanishes at $T\to\infty$ and this
ensures the specific heat approaches exactly its classical value at $T\to\infty$.
Moreover, these "arbitrarily large corrections" are sufficiently small
for realistic values of mass, box dimensions and temperature because of
the small value of the Planck constant. So, we have not found any
interesting observable prediction of the quantum mechanics in this paper
but we wanted to draw the attention to the quite interesting problem of
the high temperature limit of a quantum system. There are some
exact results in this problem. In \cite{wehrl} and references cited therein can be found
general results - inequalities between partition function and its
classical approximation -
for the systems of particles moving in space without
boundaries.

\subsubsection*{Acknowledgement}
This work was supported by the Scientific Grant Agency of Slovak Republic, project no. 1/0250/03.

\end{document}